\documentclass[11pt]{article}
\usepackage[a4paper,margin=1in]{geometry}
\usepackage{amsmath,amssymb,physics,bm}
\usepackage{authblk}
\usepackage{hyperref}
\usepackage{graphicx}

\usepackage{subcaption}
\usepackage{caption}
\usepackage{float} 

\title{Dirac Equation and Representation Dependent \\Scattering Phenomena 
}

\author[1]{Muhammad Adeel Ajaib}
\affil[1]{Penn State Abington, Abington, PA 19001, USA}

\begin{document}
\date{}
\maketitle

\begin{abstract}

We show that spin-flip probabilities emerge in the relativistic regime for scalar potentials, absent in the standard Dirac representation. We examine 1D scattering for the Dirac equation employing an alternate matrix representation introduced by the Author in an earlier study. 
We demonstrate that the transmission (\(T\)) and reflection (\(R\)) coefficients can depend on the chosen representation of the Clifford algebra despite the two representations being related by a \emph{unitary} or \emph{non-unitary} transformation. We also show that for the non-unitary case quantum interference may arise in scattering phenomena, a testable experimental signature. This representation dependence hints at hidden physics in how spinor components couple to external steps/barriers, even when the relativistic dispersion relation \(E^2=p^2+m^2\) is the same.

\end{abstract}

\section{Introduction}

The Dirac equation \cite{Dirac1928} describes spin 1/2 particles in the relativistic limit. It is the foundation of quantum field theory which has successfully described subtle physical phenomena in particle physics. The L\'evy--Leblond equation describes spin 1/2 particles in the non-relativistic limit. Another form of the L\'evy--Leblond equation was independently proposed in \cite{Ajaib2015, Ajaib2015NR} and it was shown that spin can be introduced in foundational quantum mechanics problems like the step potential and the finite barrier problem.  

This work extends the non-relativistic findings of [3] into the relativistic domain and explores both unitary and non-unitary transformations.
We show that the representation of the nilpotent matrices can be crucial in the manifestation of spin-related effects in scattering problems such as the step potential and finite barrier problems. In other words, the choice of the matrices can turn the spin related effects on or off in these scattering problems. 

The paper is organized as follows: In section \ref{sec:DiracLevy}, we discuss the Dirac equation and its non-relativistic limit. We review the known results for the scattering problem and present the nilpotent matrices corresponding to the Dirac equation. In section \ref{sec:DiracAjaib}, we discuss the representation of matrices that predict a spin--flip in the relativistic and non-relativistic limit for the step potential problem. In section \ref{sec:nonunitary} we discuss representation of the Dirac equation that do not transform to the standard Dirac equation through unitary transformations and discuss how phenomena like spin--flip and quantum interference arise in the results. We conclude in section \ref{conclusion}.

\section{The 1D Dirac Equation and Its Representation}\label{sec:DiracLevy}

The one--dimensional reduction of the Dirac equation can be written in the compact form  as
\begin{eqnarray}
-i \partial_z \psi = (i  \gamma_0 \partial_t  - i \gamma_5 m) \psi
\label{eq:Dirac}
\end{eqnarray}
In momentum space ($\psi=u(p) e^{-i p.x}=u(p) e^{-i (E t - p_z z )}$) the above equation yields,
\begin{equation}
(\gamma^{0}E - p\,I - i \gamma^{5}m)\psi = 0,
\label{eq:dirac1D}
\end{equation}
where $\gamma^{0}$ and $\gamma^{5}=i\gamma^{0}\gamma^{1}\gamma^{2}\gamma^{3}$ satisfy
\begin{equation}
(\gamma^{0})^{2}=+I,\qquad (i \gamma^{5})^{2}=-I,\qquad 
\{\gamma^{0},\gamma^{5}\}=0.
\end{equation}
This equation describes the standard 1D Dirac dynamics in a representation that isolates the
time--like and mass--like couplings in a symmetric way, convenient for comparison with the Ajaib
matrices introduced later.

Squaring Eq.~\eqref{eq:dirac1D} gives
\[
(\gamma^{0}E - p - \gamma^{5}m)^{2}
= (E^{2}-p^{2}-m^{2})I,
\]
yielding the familiar relativistic dispersion relation \(E^{2}=p^{2}+m^{2}\).

\subsection*{Non--Relativistic Limit and the \texorpdfstring{$\eta$}{eta} Representation}

As shown in Ref.~\cite{Ajaib2015NR} (see also \cite{Ajaib2015}), the non-relativistic limit of the Dirac equation yields the following form for nilpotent matrices $\eta$ defined by
\[
\eta_D = \frac{1}{\sqrt{2}}(\gamma^{0}+i\gamma^{5}), \qquad
\eta_D^{2}=0,\qquad (\eta_D^{\dagger})^{2}=0,\qquad \{\eta_D,\eta_D^{\dagger}\}=2I.
\]
Similarly, the representation introduced by L\'evy--Leblond is given by \cite{LevyLeblond:1967zz}:
\[
\eta_L = \sqrt{2} 
\begin{pmatrix}
0 & 0 \\[4pt]
1 & 0
\end{pmatrix}
\]
These representations correspond to the \emph{Dirac/L\'evy--Leblond limit} of the $\eta$--formalism
discussed in \cite{Ajaib2015}.  It reproduces both the Dirac equation and its non--relativistic
Schr\"odinger limit, but it represents a \emph{different realization of the $\eta$--algebra}
than that used in the Ajaib representation introduced in the following section.

The crucial difference lies in how the two representations treat spin coupling.  
In the \textbf{Dirac} $(\gamma^{0},\gamma^{5})$ representation, the scalar potential couples only
to the charge density, so neither the relativistic equation nor its non--relativistic limit produces
spin--flip transitions for scalar potentials.
The resulting equation is spin--diagonal, and the transmission and reflection
coefficients for a step potential preserve spin orientation.

\begin{figure}
\centering
\includegraphics[scale=.4]{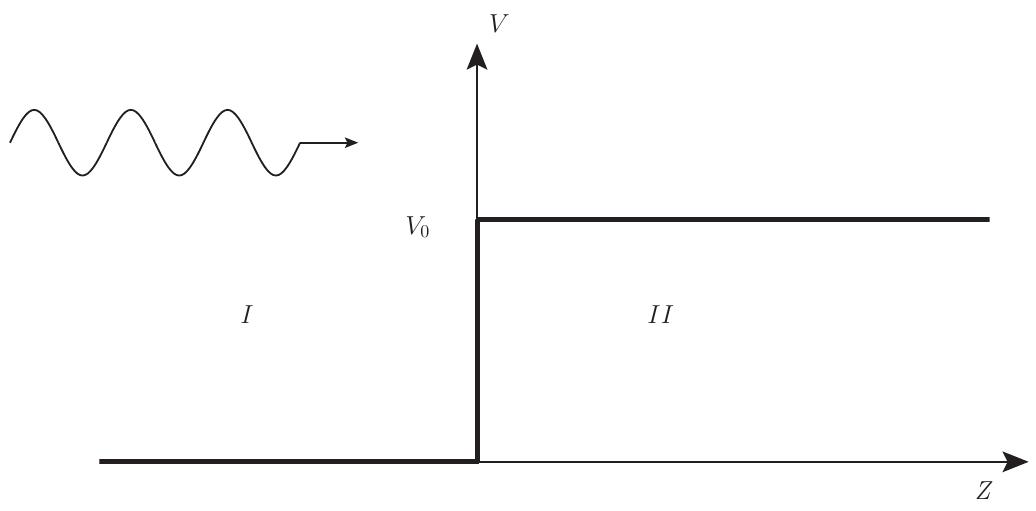}
\caption{Particle incident on a step potential.}
\label{fig:step}
\end{figure}

\begin{figure}[H]
    \centering
    \begin{subfigure}[b]{0.45\textwidth}
        \centering
        \includegraphics[width=\textwidth]{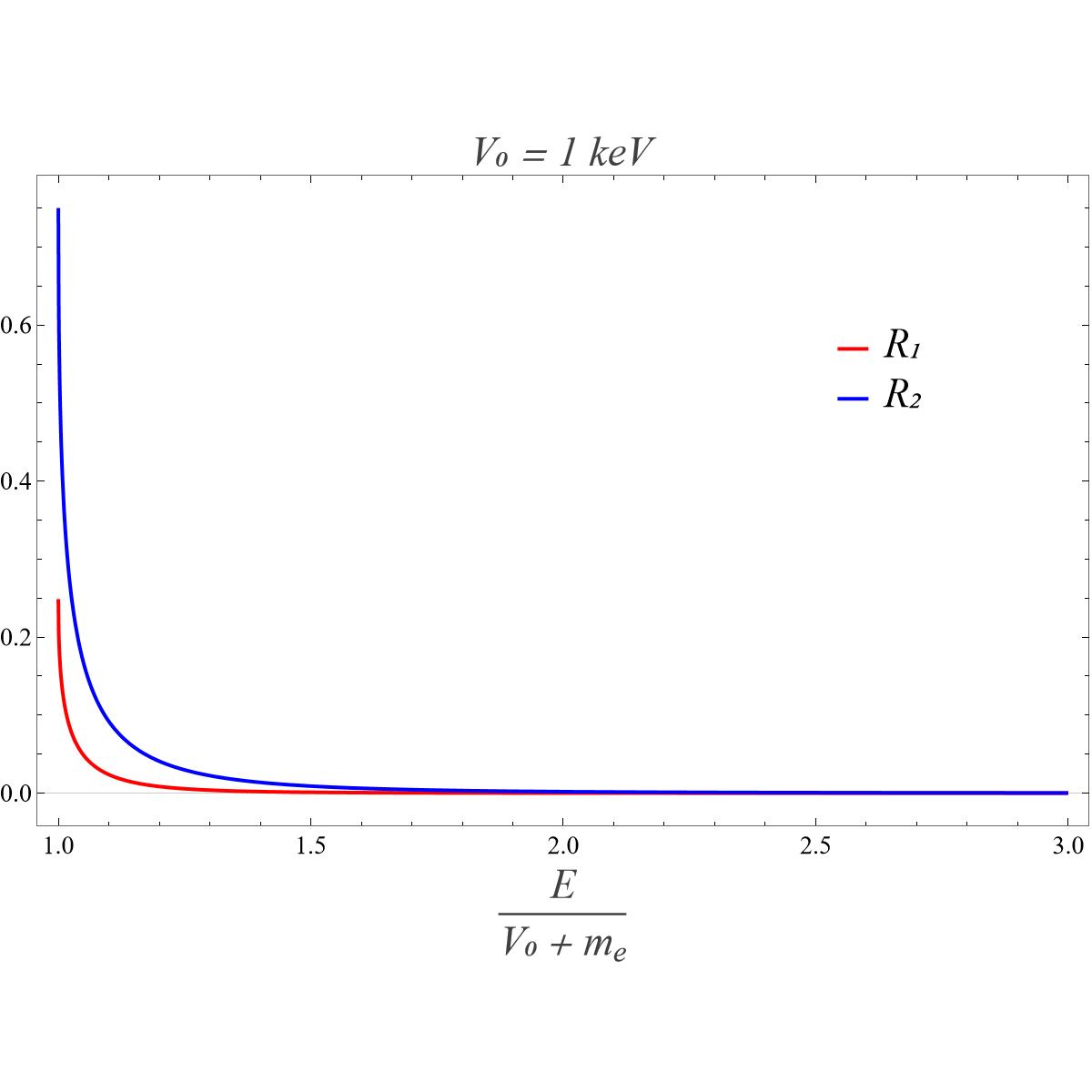}
        \caption{Spin up (red) and down (blue) Reflection coefficients}
        \label{fig:transmission}
    \end{subfigure}
    \hfill
    \begin{subfigure}[b]{0.45\textwidth}
        \centering
        \includegraphics[width=\textwidth]{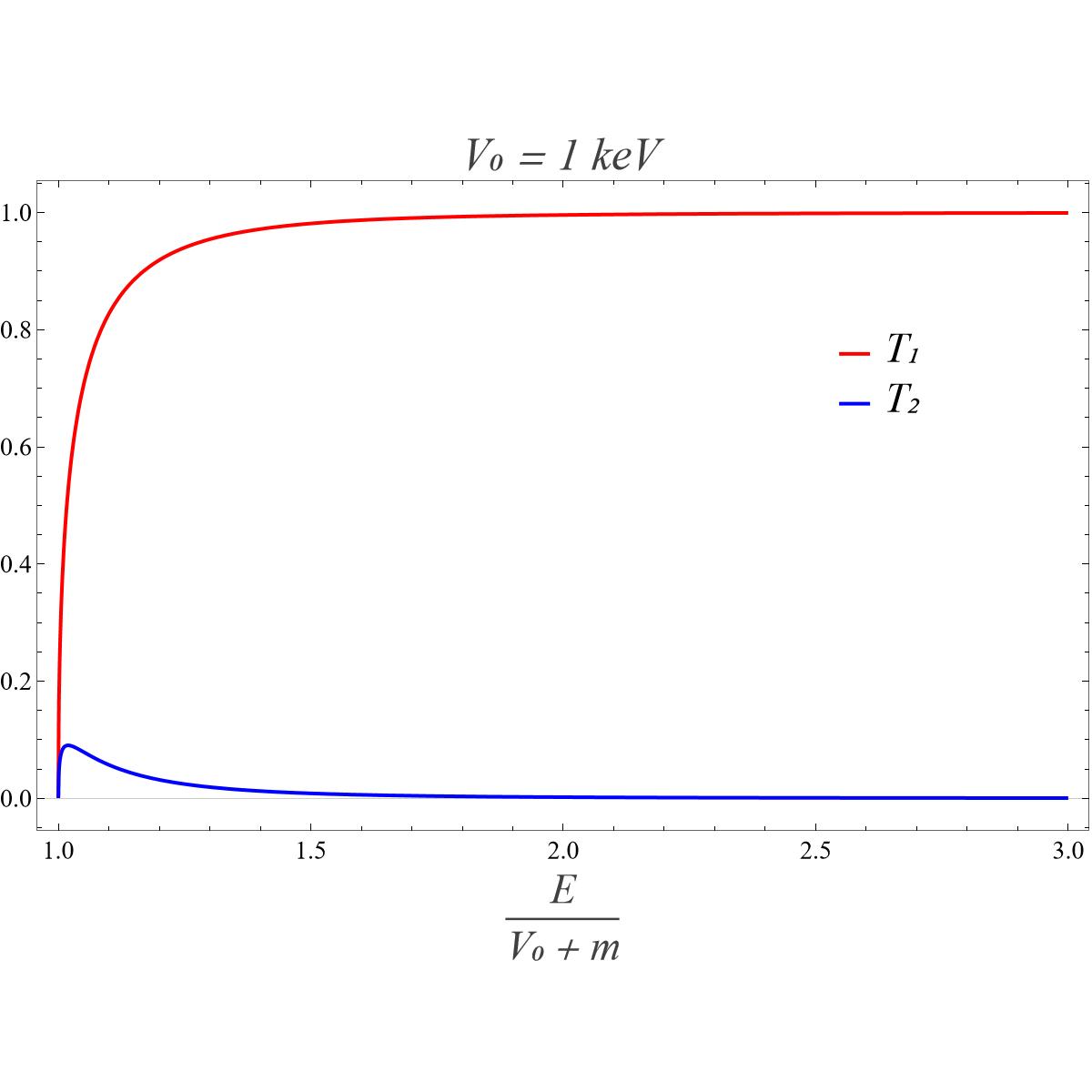}
        \caption{Spin up (red) and down (blue) Transmission coefficients}
        \label{fig:reflection}
    \end{subfigure}
    \caption{Comparison of transmission and reflection coefficients for an electron incident on a potential barrier with $V_0= 1 \ keV$. We can see that even in the relativistic case there is a non-zero probability for the electron to flip its spin. }
    \label{fig:TRplots}
\end{figure}

\section{Dirac's Equation with Ajaib's Representation}\label{sec:DiracAjaib}
A first-order equation was presented in reference \cite{Ajaib2015} that underlies the Schr\"odinger equation by employing nilpotent matrices \(\eta\) satisfying
\begin{eqnarray}
-i \partial_z \psi = (i  \eta \partial_t  + \eta^\dagger m) \psi
\label{mse-eq}
\end{eqnarray}
where $\eta$ is a nilpotent matrix given by,
\begin{eqnarray}
\eta &=& 
 -\frac{i}{\sqrt{2}}
\left( \gamma_0 \gamma_1 \gamma_5 + \gamma_2
\right)
=
\frac{-i}{\sqrt{2}}\left(
\begin{array}{cc}
  \sigma _1 &  \sigma _2 \\
 -\sigma _2 &  \sigma _1
\end{array}
\right)
\label{eq:matrix-rep1}
\end{eqnarray}
and 
\begin{equation}
\eta^2=(\eta^\dagger)^2=0,\qquad \{\eta,\eta^\dagger\}=2\mathbb{I},
\label{eq:nilpotent}
\end{equation}

The 1D step potential problem was analyzed and it was shown that spin effects emerge in the non-relativistic limit~\cite{Ajaib2015}. The step potential and the finite potential barrier problem were solved and it was shown that the particle's spin can be included in these problems. Moreover, the sum of the spin up/down transmission and reflection coefficients yields the known quantum mechanical results. It was also shown that quantized energy
levels for the hydrogen atom are obtained and the result is consistent with non-relativistic quantum mechanics \cite{Ajaib2017cpl}. The 2D scattering of electrons from a finite potential barrier was also analyzed and it was shown that spin related effects also exhibit in these scattering problems \cite{Ajaib2017zfk}.

Following the standard approach \cite{Ajaib2015} we can show that Equation (\ref{mse-eq}) is the non-relativistic limit of the following equation:
\begin{eqnarray}
-i \partial_z \psi = (i  \chi_1 \partial_t  + \chi_2 m) \psi
\label{eq:rel-ver}
\end{eqnarray}
where,
\begin{align*} 
\chi_1  = \frac{\eta+\eta^\dagger}{\sqrt{2}}= -i \gamma_2,& \qquad
\chi_2  = \frac{\eta-\eta^\dagger}{\sqrt{2}} = -i \gamma_0 \gamma_1 \gamma_5,\qquad \\ \nonumber
\\ 
\chi_1^2 = +\mathbb{I},&\quad \chi_2^2=-\mathbb{I},\quad \\ \nonumber \chi_1^\dagger=\chi_1,&\quad \chi_2^\dagger=-\chi_2\  \\ \nonumber
\{\chi_1,\chi_2\} & =  0.
\label{eq:X1X2}
\end{align*} 

The dispersion relation for a particle described by equation (\ref{eq:rel-ver}) is that of a relativistic particle (\(E^2=p^2+m^2\)). Unlike the \emph{Dirac/L\'evy--Leblond} equation, this relativistic equation predicts a spin-flip in the step potential problem also. 
In momentum space, the plane-wave form of Equation (\ref{eq:rel-ver}) is given by,
\[
(\chi_1 E - p_z - \chi_2 m) u = 0,
\]
The probability density and current density is given by,
\begin{eqnarray}
\rho_A = \psi^\dagger\psi,\quad j_A = \psi^\dagger \chi_1 \psi
\end{eqnarray}

We consider the step potential problem shown in Fig. \ref{fig:step}. Following the standard continuity approach, the spin-resolved transmission and reflection coefficients are obtained as follows,
\begin{align}
T_{\uparrow} &= 
\frac{\widetilde{\Delta}(E,V_0)^2}{E^2 V_0^2},
&
R_{\uparrow} &= 
\frac{m^4 V_0^2}{E^2\,\Delta(E,V_0)^2},
\label{eq:TupRup}\\[0.8em]
T_{\downarrow} &= 
\frac{(E^2-m^2) m^2 V_0^2}
     {E^2\,\Delta(E,V_0)^2},
&
R_{\downarrow} &= 
\frac{(E^2-m^2) m^2 V_0^2}
     {E^2\,\Delta(E,V_0)^2}.
\label{eq:TdownRdown}
\end{align}

The total transmission and reflection probabilities follow from

\begin{align}
T &= T_{\uparrow} + T_{\downarrow} = \frac{2(E^2-m^2)(E-V_0)}{E\,\Delta(E,V_0)}, \label{eq:total_transmission}\\ 
\qquad
R &= R_{\uparrow} + R_{\downarrow} = \frac{m^2 V_0^2}{\Delta(E,V_0)^2}\\ 
\label{eq:total_reflection}\qquad
T&+R =1
\end{align}
where,
\begin{align}
\Delta(E,V_0) &=
E^2 - m^2 
+ \sqrt{(E^2 - m^2)\!\left[(E - V_0)^2 - m^2\right]}
- E V_0, 
\label{eq:Delta}\\[0.6em]
\widetilde{\Delta}(E,V_0) &=
E^2 - m^2 
- \sqrt{(E^2 - m^2)\!\left[(E - V_0)^2 - m^2\right]}
+ E V_0.
\label{eq:DeltaTilde}
\end{align}
These expressions correspond to the scattering of a relativistic particle governed by the 
Dirac equation in Ajaib’s representation, where \(E\) is the total energy, \(m\) the rest mass,
and \(V_0\) the step potential height. Note that the total transmission and reflection coefficients given in equations (\ref{eq:total_transmission}) and (\ref{eq:total_reflection}) are also obtained for the Dirac equation but with $T_\downarrow =0$ and $R_\downarrow=0$. The spin-flip originates from the non-diagonal structure of $\chi_2$, which couples upper and lower components differently across the potential interface.

We now consider the asymptotic behavior of the transmission and reflection coefficients in the limit \( V_0 \to \infty \).
We obtain the following forms for the transmission and reflection coefficients:
\begin{align}
\lim_{V_0 \to \infty} T &= \frac{2(E^2 - m^2)(\sqrt{E^2 - m^2} + E)}{ m^2}, \\[0.5em]
\lim_{V_0 \to \infty} R &= \frac{(\sqrt{E^2 - m^2} + E)^2}{m^2}.
\end{align}

These expressions indicate that the Klein paradox persists even within this representation. In particular, the reflection coefficient can exceed unity in the high-potential barrier limit, while transmission also remains nonzero, consistent with Klein tunneling. Thus, despite using an alternate but unitarily equivalent matrix representation, the paradoxical behavior intrinsic to relativistic quantum mechanics is preserved.

\subsection*{Unitarity and Physical Distinctions}

A similarity \(S\) exists such that 
\(
\chi_1 = S\,\gamma^0 S^{-1},\quad \chi_2 = S\,(i\gamma^5) S^{-1},
\)
but with \(S^\dagger S = \mathbb{I}\).  
The two sets of matrices \(\{\gamma^\mu\}\) and \(\{X_\mu\}\) satisfy the same Clifford algebra and are related by a unitary transformation:
\begin{equation}
\chi_\mu = S\,\gamma_\mu\,S^{-1},\qquad S^\dagger S = \mathbb{I}.
\end{equation}

It is important to emphasize that the modified relativistic equation proposed by the author is related to the Dirac equation via a \emph{unitary} transformation. This implies that the two theories are equivalent at the level of the free particle Hamiltonian and share the same dispersion relation \( E^2 = p^2 + m^2 \).

However, despite this equivalence, our calculations show that the transmission and reflection coefficients differ between the two representations when analyzing scattering from a step potential. The key distinction lies in the probability current operator: in the standard Dirac representation, $j_D = \psi^\dagger \gamma^0 \psi$, while in the Ajaib representation, $j_A = \psi^\dagger \chi_1 \psi$. When solving the step potential problem, boundary conditions require continuity of both $\psi$ and $j$ at the interface. Since $\chi_1 \neq \gamma^0$, these representations lead to algebraically different spinor matching conditions, even though $\psi_A = S\psi_D$.

This suggests that while the bulk dynamics remain invariant, interface phenomena such as scattering and spin-flip may encode new physics that is sensitive to the choice of spinor representation. It has been shown that self-adjoint extensions of formally equivalent Hamiltonians can yield inequivalent scattering matrices \cite{Bonneau:1999zq, ReedSimonII}.

The observed differences in transmission and reflection coefficients suggest the possibility of experimentally testing the predictions of this representation-dependent formalism. In principle, electron scattering from sharp step potentials or junctions with tunable barrier heights could reveal signatures of this formalism. We encourage the development of dedicated experimental setups to explore these possibilities.

\section{Non-Unitary Representation and Quantum Interference}
\label{sec:nonunitary}

In this section, we introduce an alternative representation of the Dirac equation that is \emph{not} unitarily related to the standard form. This representation exhibits rich multi-channel quantum interference phenomena that are fundamentally absent in the standard Dirac theory. We begin by presenting the non-relativistic nilpotent matrices $\eta_1$ and $\eta_2$, show their relationship to the relativistic $\xi_1$ and $\xi_2$ matrices, and demonstrate that the resulting scattering dynamics reveal momentum-degenerate channels with distinct spin structures that interfere coherently.

\subsection*{The Non-Relativistic \texorpdfstring{$\eta_1$, $\eta_2$}{y1, y2} Matrices}

We define the $4 \times 4$ nilpotent matrices $\eta_1$ and $\eta_2$ as follows:

\begin{align}
\eta_1 &= \frac{i}{\sqrt{2}} \ \gamma_0 (\gamma_5 + \gamma_2) 
\\ 
\eta_2 &= i \sqrt{2} \ (\gamma_2 + \gamma_0 )\gamma_5 
\label{eq:y_matrices}
\end{align}

These matrices satisfy the following algebraic properties:
\begin{align}
\eta_1^2 &= \eta_2^2 = 0, \label{eq:y_nilpotent}\\
\{\eta_1, \eta_2\} &= 2\mathbb{I}, \label{eq:y_anticomm}
\end{align}

The nilpotency conditions  and the anticommutation relations define a generalized Clifford-like algebra structure. These matrices generate a four-dimensional spinor representation distinct from both the standard Dirac and L\'evy-Leblond forms.

A first-order equation incorporating these matrices can be written as:
\begin{equation}
-i \partial_z \psi = (i \eta_1 \partial_t + \eta_2) \psi.
\label{eq:y_equation}
\end{equation}

In momentum space, where $\psi = u(p) e^{-i(Et - p_z z)}$, this becomes:
\begin{equation}
p_z u = (\eta_1 E + \eta_2 m) u.
\label{eq:y_momentum}
\end{equation}

Squaring Eq.~\eqref{eq:y_momentum} and using the algebraic properties~\eqref{eq:y_nilpotent}--\eqref{eq:y_anticomm}, we obtain:
\begin{equation}
p_z^2 = (\eta_1 E + \eta_2 m)^2 =  \{\eta_1, \eta_2\} E m = 2 E m,
\label{eq:y_dispersion}
\end{equation}
which is the \emph{non-relativistic dispersion relation}. This demonstrates that the $\eta$-matrices naturally encode non-relativistic particle dynamics.

\subsection*{The Relativistic \texorpdfstring{$\xi_1$, $\xi_2$}{z1, z2} Matrices}

As discussed in Section \ref{sec:DiracAjaib}, the relativistic matrices  $\xi_1$ and $\xi_2$ can be obtained from $\eta_1$ and $\eta_2$ as follows:
\begin{equation}
\xi_1 = \frac{\eta_1 + \eta_2}{\sqrt{2}}, \qquad \xi_2 = \frac{\eta_1 - \eta_2}{\sqrt{2}}.
\label{eq:z_definition}
\end{equation}

From the properties of $\eta_1$ and $\eta_2$, we can derive the algebraic structure of $\xi_1$ and $\xi_2$:
\begin{align}
\xi_1^2 &=  +\mathbb{I}, \label{eq:z1_square}\\
\xi_2^2 &=  -\mathbb{I}, \label{eq:z2_square}\\
\{\xi_1, \xi_2\} &= 0. \label{eq:z_anticomm}
\end{align}

The first-order relativistic equation using $\xi_1$ and $\xi_2$ is given by,
\begin{equation}
-i \partial_z \psi = (i \xi_1 \partial_t + \xi_2 m) \psi.
\label{eq:z_equation}
\end{equation}

In momentum space:
\begin{equation}
p_z u = (\xi_1 E + \xi_2 m) u.
\label{eq:z_momentum}
\end{equation}

Squaring the above equation shows that the eigenvalues of $\xi_1 E + \xi_2 m$ satisfy:
\begin{equation}
p_z^2 = E^2 - m^2,
\label{eq:z_dispersion}
\end{equation}
which is the \textbf{relativistic dispersion relation} $E^2 = p^2 + m^2$.


The transformation from the standard Dirac matrices $(\gamma^0, \gamma^5)$ to $(\xi_1, \xi_2)$ is a similarity but not a unitary transformation:
\begin{equation}
\xi_1 = U \gamma^0 U^\dagger, \qquad \xi_2 = U (i\gamma^5) U^\dagger.
\end{equation}
which induces a metric, $g = U^\dagger U \neq 1$. This metric defines a modified inner product $\langle \psi | \phi \rangle_g = \psi^\dagger g \phi$ in the $\{\xi_1, \xi_2\}$ representation. Natural symmetries preserving the Clifford algebra must satisfy $U^\dagger g U = g$ ($g$-unitarity), which differs from standard unitarity $U^\dagger U = I$.
Note that the Hamiltonian in this representation will be \emph{pseudo-Hermitian} with respect to $g$, that is, $H^\dagger  g = g H$, with real eigenvalues and unitary time evolution \cite{Mostafazadeh:2001jk}. 
This non-unitarity has notable consequences: while the bulk dispersion relation remains the same, the \emph{boundary physics} and \emph{scattering properties} can differ dramatically. In particular, the probability current  operator is hermitian, given by:
\begin{equation}
j_z = i \psi^\dagger (\gamma_2 \gamma_3 + \gamma_2) \psi.
\label{eq:current_z}
\end{equation}
It differs from the standard Dirac current $j_D = \psi^\dagger \gamma^0 \psi$, leading to distinct boundary matching conditions and hence different transmission and reflection amplitudes.

\subsection*{Step Potential, Spin flip and Quantum Interference}

We now analyze the one-dimensional step potential problem (Fig.~\ref{fig:step}) with these matrices. In momentum space, the equation in each region is given by:
\begin{align}
p_z &= \xi_1 E + \xi_2 m, \quad \quad \text{Region } I \label{eq:H_I}\\
p_z &= \xi_1 (E - V_0) + \xi_2 m  \quad \quad \text{Region } II \label{eq:H_II}
\end{align}

\begin{figure}[t]
\centering
\begin{subfigure}[b]{0.48\textwidth}
    \centering
    \includegraphics[width=\textwidth]{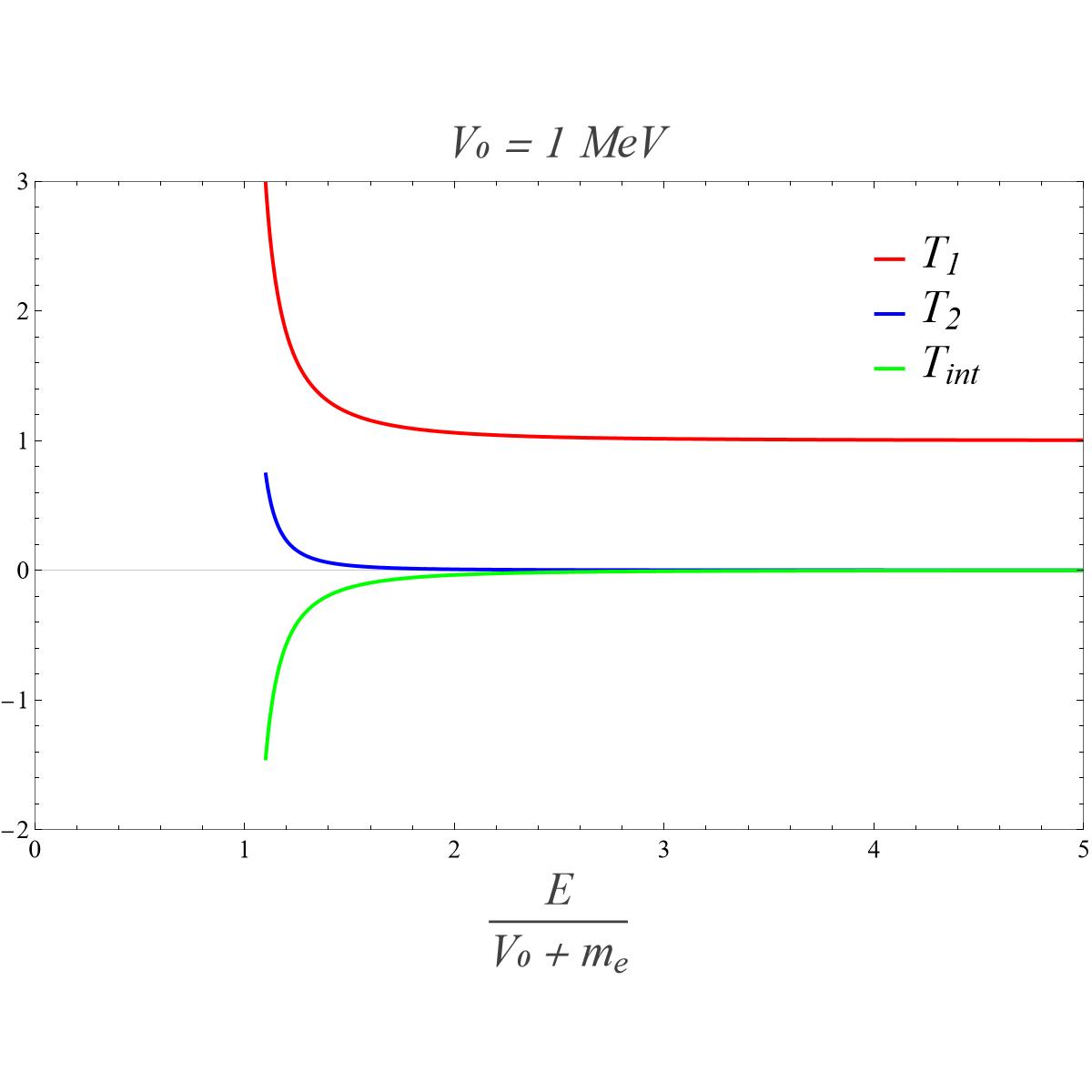}
    \caption{}
    \label{fig:transmission_components}
\end{subfigure}
\hfill
\begin{subfigure}[b]{0.48\textwidth}
    \centering
    \includegraphics[width=\textwidth]{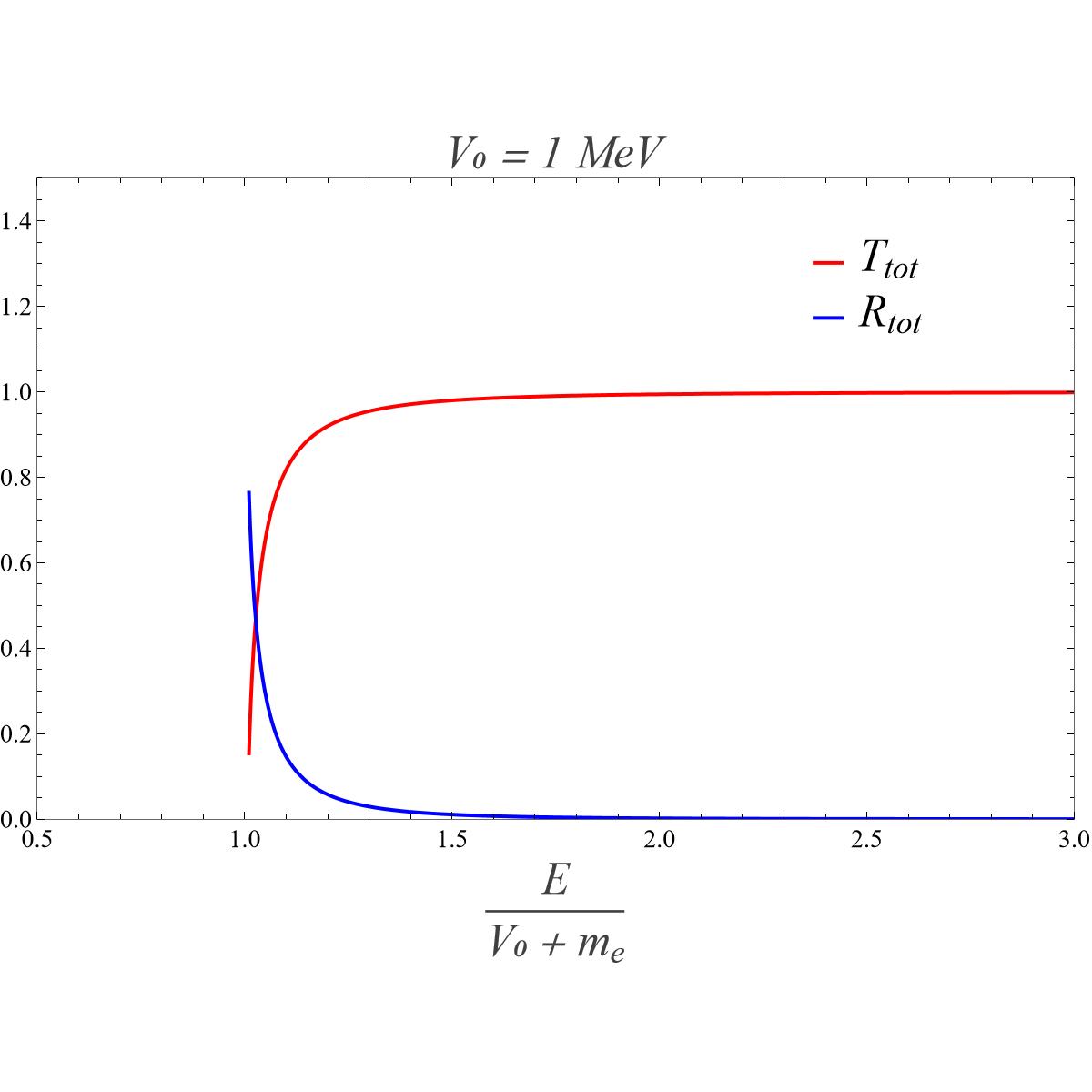}
    \caption{}
    \label{fig:transmission_conservation}
\end{subfigure}
\caption{Energy dependence of transmission and reflection coefficients for a spin-up incident electron encountering a step potential in the $\{\xi_1, \xi_2\}$ representation for the $E > V_0 + m$ case. 
(a)~Decomposition of total transmission into amplitude components: $T_1$ (red, diagonal contribution $|t_{\uparrow\uparrow}|^2$), $T_2$ (blue, off-diagonal contribution $|t_{\uparrow\downarrow}|^2$), and $T_{\text{int}}$ (green, quantum interference term $2\text{Re}(t^*_{\uparrow\uparrow} t_{\uparrow\downarrow})$). The large individual magnitudes and substantial negative interference produce the physical total transmission $T_{\text{tot}} = T_1 + T_2 + T_{\text{int}} \le 1$, reflecting strong quantum correlations between spin channels. 
(b)~Total transmission $T_{\text{tot}}$ (red) and reflection $R_{\text{tot}}$ (blue) demonstrating probability conservation $T_{\text{tot}} + R_{\text{tot}} = 1$ (black dashed line). The transition from reflection-dominated to transmission-dominated behavior exhibits Klein tunneling, with near-perfect transmission at high energies.}
\label{fig:transmission_twoPanel}
\end{figure}

Figure~\ref{fig:transmission_twoPanel} presents the energy-dependent transmission and reflection coefficients for a relativistic spin-up incident particle incident towards a step potential in the $\{\xi_1, \xi_2\}$ representation. 
Panel~(a) displays the decomposition of the total transmission probability into its constituent amplitude components. The total transmission $T_{\text{tot}}$ can be expressed as
\begin{equation}
T_{\text{tot}} = T_1 + T_2 + T_{\text{int}},
\label{eq:transmission_decomposition}
\end{equation}
where $T_1$ (spin up) and $T_2$ (spin down) represent the squared magnitudes of the diagonal and off-diagonal transmission amplitudes, respectively, while $T_{\text{int}}$ captures the quantum interference between these channels:
\begin{equation}
T_1 = |t_{\uparrow\uparrow}|^2, \quad T_2 = |t_{\uparrow\downarrow}|^2, \quad T_{\text{int}} = 2\text{Re}(t^*_{\uparrow\uparrow} t_{\uparrow\downarrow}).
\end{equation}

Several interesting features emerge from this decomposition. First, the diagonal component $T_1$ (red curve) exhibits values significantly exceeding unity at low energies near threshold, reaching approximately $T_1 \approx 3$ at $E/(V_0+m) \approx 1.1$. This behavior is consistent for an amplitude contribution when interference terms are present. The large magnitude of $T_1$ reflects the strong coupling between spin states induced by the off-diagonal structure of the Hamiltonian in the $\{\xi_1, \xi_2\}$ basis.
Furthermore, the off-diagonal component $T_2$ (blue curve) shows a pronounced peak near threshold before rapidly decaying at higher energies. This peak structure, reaching approximately unity at $E/(V_0+m) \approx 1.2$, indicates enhanced spin-flip scattering in the near-threshold regime where the particle momentum becomes comparable to the characteristic scales set by the potential step and mass.

The interference term $T_{\text{int}}$ (green curve) is large and \emph{negative} close to the threshold value. This substantial negative interference between the diagonal and off-diagonal channels is essential for ensuring that the total transmission remains physical ($T_{\text{tot}} \le 1$). The large negative interference term ($T_{\text{int}} \approx -3$ near threshold) has 
a direct physical interpretation: the transmission amplitudes $t_{\uparrow\uparrow}$ and $t_{\uparrow\downarrow}$ have nearly opposite phases, leading to destructive 
quantum interference. This destructive interference is essential for 
maintaining $T_{\text{tot}} \le 1$.
At high energies ($E/(V_0+m) \gtrsim 2$), all three components asymptotically approach constant values. In this regime, $T_1 \to 1$, $T_2 \to 0$, and $T_{\text{int}} \to 0$, yielding $T_{\text{tot}} \to 1$ as expected for Klein tunneling. 
Panel~(b) verifies probability conservation: $T_{\text{tot}} + R_{\text{tot}} = 1$ throughout the energy range. The transition from reflection-dominated ($R_{\text{tot}} \approx 0.9$) to transmission-dominated ($T_{\text{tot}} \to 1$) behavior exhibits the characteristic Klein tunneling of relativistic particles.

\section{Conclusions}\label{conclusion}

We have demonstrated that one-dimensional Dirac-like scattering problems can exhibit representation dependence for underlying theories related by a unitary or non-unitary transformation. This study is an extension of similar results shown by the Author for non-relativistic energies in an earlier study \cite{Ajaib2015}. In both the relativistic regime and the non-relativistic limit, the representation presented in our analysis predicts spin-flip transitions and quantum interference for scalar step potentials—a phenomenon entirely absent in the standard Dirac representation, where scalar potentials couple only to charge density and preserve spin orientation.

This sensitivity of interface physics to representation choice has 
implications for quantum information processing, spintronics, and 
spin-based quantum devices. Future work could explore representation-
dependent effects for other potential profiles, in the presence of 
magnetic fields, or in time-dependent scattering, where the interplay 
between spin and boundary conditions becomes even more intricate.

\section{Acknowledgments}
We would like to thank Abdul Rehman and Fariha Nasir for useful discussions.
\vspace{.5cm}

The mathematica code for the analysis in the paper is available on github: \url{https://github.com/dradeelajaib/relativistic-spin}

\end{document}